\documentclass[12pt]{article}
\usepackage{amssymb,epsf}
\newcommand{\Fio}[3]{{\itshape #3~#1.\nolinebreak#2.}}
\newcommand{\Fi}[2]{{\itshape #2~#1.\nolinebreak}}
\def\Journal#1#2#3#4{// {#1} {\bf #2} (#4) #3}
\def\NPB{Nucl. Phys.}
\def\PLB{Phys. Lett.}
\def\PRL{Phys. Rev. Lett.}

\def\PRD{Phys. Rev.}
\def\ZPC{Z. Phys.}
\def\MPLA{Mod. Phys. Lett.}
\def\JMPA{J. Mod. Phys.}
\newcommand{\op}[1]{\mathop{\rm #1}\nolimits}
\newcommand{\up}[1]{\mathrm{#1}}
\newcommand{\be}{\begin{equation}}
\newcommand{\ee}[1]{\label{#1}\end{equation}}
\newcommand{\bear}{\begin{equation}\begin{array}}
\newcommand{\eear}[1]{\end{array}\label{#1}\end{equation}}
\newcommand{\ol}{\overline}
\newcommand{\ds}{\displaystyle}
\newcommand{\vep}{\varepsilon}

\begin{document} \large
\begin{center}
{\LARGE
 Free Energy of an $SU(2)$ Model of (2+1)-dimensional QCD in
 the Constant Condensate Background \\
} \bigskip
{\large\slshape V.\,Ch. Zhukovskii, V.\,V. Khudyakov}\bigskip \\
{\small\itshape Department of Theoretical Physics, \\
 Physical Faculty, Moscow State University, \\
 117234, Moscow, Russia\\
E-mail: th180@phys.msu.su
 }
\end{center}
\medskip

\begin{abstract} \normalsize
 Gluon and quark contributions to the thermodynamic potential (free energy)
 of a (2+1)-dimensional QCD model at finite temperature in the background of
 a constant homogeneous chromomagnetic field $H$ combined with
 $A_0$ condensate is calculated. The role of the tachyonic mode
 in the gluon energy spectrum is discussed. A possibility of
 the free energy global minimum generation at nonzero values of $H$ and
 $A_0$ condensates is investigated.
\end{abstract}

\section*{Introduction}

Quantum field theories in low dimensional space-time have recently excited a
considerable interest~\cite{n33,a16,a0} due to a close relation with
their (3+1)-dimensional
analogues~\cite{red0,VshZhu,GusMir},  as well as to a
possibility of explaining certain natural phenomena with their help. For
example, the method of dimensional reduction~\cite{red1} allows
one to study the QCD$_{3+1}$ quark-gluon plasma in the high temperature limit
above the critical temperature of the confinement--deconfinement phase
transition. As another example, we mention the
two-dimensional model of
electrons interacting with external electromagnetic fields that was employed
in the explanation of the quantum Hall effect~\cite{hall,orange}.

In the last years, much attention has been paid to investigation of the
effective potential in the background field both at zero and at high
temperatures in the one loop
approximation~\cite{td1,td2,Reinh,Weiss,td2++,Skal}, and also with
account of higher loop contributions~\cite{tdm1,tdm2,sklz,skalnew,gies}. The
assumed $A_0$-condensate formation in QCD (see~\cite{td1}) leads to important
physical consequences, such as spontaneous breaking of global gauge symmetry,
stabilization of the effective potential, elimination of the infrared
divergence etc. New results for the two-loop gauge field effective potential
in the presence of an external chromomagnetic field have been obtained
recently in~\cite{skalnew}. At the same time, a problem of gauge dependence
of the results arises~\cite{sklz} in the case of nonzero $A_0$ potential,
when higher orders of the perturbation expansion are taken into account.

In the present paper, we study the one-loop gauge field effective potential
(thermodynamic potential) in the (2+1)-dimensional QCD model in the presence
of a $A_0$ potential and a chromomagnetic field at finite temperatures. In
the first section, some basic formulas that describe quantum field  systems
at finite temperatures are reminded, and the relation between the effective
potential and the energy spectrum of one-particle excitations is
demonstrated. In the second section, the gluon field contribution to the
free energy density is calculated in the (2+1)-dimensional model of QCD.  The
tachyonic mode in the energy spectrum of gluons is demonstrated to lead to a
nonanalytic dependence of the thermodynamic potential on the condensate field
strength. A possible way of removing this nonanalytic behavior is presented.
A series of nontrivial minima of the free energy, as well as phases of
confinement and deconfinement are shown to exist when temperatures are below
the critical value. In the third section, the quark contribution to free
energy density is obtained. Calculations show that, in contrast to the gluon
sector, its minimum exists in the absence of the condensate fields.

\section{Energy spectrum of excitations and free energy}

Consider a QCD$_{2+1}$ model with a gluon gauge field $A^a_\mu$ in the
adjoint representation of the $SU(2)$ color group. It can be written as the
sum $A^a_\mu=\ol{A}^a_\mu+a^a_\mu,$ where $\ol{A}^a_\mu$ is the background
potential and $a^a_\mu$ are the gluon quantum fluctuations. The
gauge field Lagrangian in the Lorentz gauge in the Euclidean space-time
has the form
\be
  {\cal L}_\up{g}= \frac{1}{4}(F^a_{\mu\nu})^2+
  \frac{1}{2}(\ol{D}^{ab}_\mu a^{b}_{\mu})^2+
  \ol{\chi}^a(\ol{D}^2)^{ab}\chi^b \;,
\ee{r1}
where
$\ol{D}^{ab}_\mu=\delta^{ab}\partial_\mu-gf^{abc}\ol{A}^c_\mu$
is the covariant derivative, $\chi, \ol{\chi}$ are ghost fields,
and $(\ol{D}^2)^{ab} =\ol{D}^{ac}_\mu \ol{D}^{cb}_{\mu}.$ In passing to
the Euclidean space-time, we retained the subscript $0$ to denote
those components of vectors that are time-like, so that
$x_0=it$ and $\mu,\nu = 0,1,2.$ The following representation of the
Clifford algebra can be chosen in the 3-dimensional space-time:
\be
  \gamma^0=\sigma^3, \gamma^1=i\sigma^1, \gamma^2=i\sigma^2,
\ee{r2}
where $\sigma^1, \sigma^2, \sigma^3$ are Pauli matrices. Then the Lagrangian
for quarks with $N_\up{f}$ flavors takes the form
\be
  {\cal L}_\up{q}=\sum^{N_\up{f}}_{j=1}\ol{\psi}_j\left[\gamma_\mu
  (\partial_\mu- ig\frac{\lambda_a}{2}A^a_\mu)+m_j\right]\psi_j \;.
\ee{r3}
Here $\lambda_a$ matrices are the color group generators.
The generating functional
\be
  Z[\,\ol{A},j,\eta,\ol{\eta}]=\int\! da^a_\mu
  d\ol{\chi}d\chi d\ol{\psi}d\psi \exp\left[ -\int\! d^3x({\cal L}
  +j^a_\mu a^{a}_{\mu}+ \ol{\psi}\eta+\ol{\eta}\psi)\right],
\ee{r4}
where
${\cal L}={\cal L}_\up{g}+{\cal L}_\up{q}$ is the QCD Lagrangian, can be
calculated in the one loop approximation. To this end, it is sufficient
to retain in ${\cal L}_\up{g}$ only the terms quadratic in gluon
fluctuations $a^a_\mu.$ Namely,
\bear{c}
  \ds {\cal L}_\up{g}={\cal L}_\up{g}^{(0)}+{\cal L}_\up{g}^{(2)} \,,
  \quad {\cal L}_\up{g}^{(0)}=\frac{1}{4}(\ol{F}_{\mu\nu}^a)^2
  \,, \vspace{3mm}\\ \ds {\cal L}_\up{g}^{(2)}=-\frac{1}{2}a^{a}_{\mu}
  \left[ (\ol{D}^2)^{ab}\delta_{\mu\nu}+
  2g\ol{F}_{\mu\nu}^c f^{abc}\right]a^{b}_{\nu} \;,
\eear{r5}
where $\ol{F}_{\mu\nu}^a=\partial_\mu \ol{A}_\nu^a
-\partial_\nu \ol{A}_\mu^a+gf^{abc}\ol{A}_\mu^b \ol{A}_\nu^c.$
In this case the path integrals in (\ref{r4}) take the Gaussian form
and can be easily calculated:
\bear{rl}
  Z[\,\ol{A}\,]=&\exp\ds\left\{-\frac{1}{4}\int\!d^3x
  (\ol{F}^a_{\mu\nu})^2 \right\} \left[ \det(-
  \ol{D}^2\delta_{\mu\nu} -2g\ol{F}_{\mu\nu}f) \right]^{-
  \frac{1}{2}} \times \vspace{1.5mm}\\
  & \ds \times \det(-\ol{D}^2)\prod^{N_\up{f}}_{j=1}\det\left[
  \gamma_\mu(\partial_\mu-
  ig\frac{\lambda_a}{2}\ol{A}^a_\mu)+m_j\right].
\eear{r6}
Using $Z=\exp(W_\up{E}),$ one obtains the effective Euclidean action
\be
  W_\up{E}=-\frac{1}{2}\int\frac{dq_0}{2\pi}\sum_r\ln(q_0^2+(\vep^\up{g}_r)^2)
  +\sum_{j=1}^{N_\up{f}}\int\!\frac{dp_0}{2\pi}\sum_k\ln(p_0^2
  +(\vep^\up{q}_{jk})^2).
\ee{r7}
 We have taken into account here that contributions of the
gauge field longitudinal components and those of the ghost fields
cancel each other out. Summation over the quantum numbers $r$ and $k$
of the gluon $\vep^\up{g}_r$ and quark $\vep^\up{q}_{jk}$ energy spectra
should include only physical degrees of freedom. Formally, we will
consider our field system in a three-dimensional cube with volume $L^3$
and periodic boundary conditions. In this case, the effective potential
$V$ is determined as follows:
\be
  V=-\frac{W_\up{E}}{L^3}.
\ee{r8}
The effective potential $V$ in the one-loop approximation can be represented
as the sum of two terms: $V=V^{(0)}+v,$ where
$V^{(0)}=(\ol{F}_{\mu\nu}^a)^2/4$ is the energy density of the background
field, $v=v^\up{g}+v^\up{q}$ is the sum of the one-loop contributions
of gluon and quark fluctuations. The limit of an infinite space-time
volume is obtained, when $L\to \infty.$ Since the explicit expression
for the effective potential (\ref{r17}) does not depend on $L,$ this
limiting procedure will not be mentioned in what follows.

The temperature $T\equiv 1/\beta$ is introduced by imposing periodicity and
antiperiodicity conditions  on the boson  and the fermion fields respectively
in variable $x_0$ with the period equal to $\beta.$ According to the well
known prescriptions, this is achieved by introducing the Matsubara
frequencies instead of $q_0,\,p_0\,$ and making the following substitutions:
$q_0\to 2\pi l/\beta- i\varkappa\mu_1$ for bosons and $p_0\to
2\pi(l+\frac{1}{2})/\beta- i\varkappa\mu_2$ for fermions, where
$\mu_1,\mu_2$ are the gluon and quark chemical potentials respectively,
$\varkappa=\pm 1$ determines the particles and antiparticles charge signs,
$l\in \mathbb{Z}$ ($\mathbb{Z}$ is the set of integers). The one-loop
effective potential, determined in (\ref{r8}), is proportional to the free
energy density
\be
  V=\frac{-T\ln(Z)}{L^2}=\frac{\Omega}{L^2}.
\ee{r10}
Upon the above substitution, the quark and gluon contributions to the
one  loop effective potential take the form
\bear{rl}
  v=&v^\up{g}+v^\up{q}=
  \ds\frac{1}{2\beta L^2}\sum_{l=-\infty}^{+\infty}\sum_{r,\,\varkappa=\pm 1}
  \ln\left[\left(\frac{2\pi l}{\beta}-i\varkappa\mu_1\right)^2
  +(\vep^\up{g}_r)^2 \right]- \vspace{2mm}\\ &
  \ds -\frac{1}{\beta L^2}\sum_{j=1}^{N_\up{f}}
  \sum_{l=-\infty}^{+\infty} \sum_{k,\,\varkappa=\pm 1}
  \ln\left[\left(\frac{2\pi (l+1/2)}{\beta}
  -i\varkappa\mu_2\right)^2+(\vep^\up{q}_{jk})^2 \right].
\eear{r11}

We introduce the background field as a superposition of the constant
homogeneous chromomagnetic field $H,$ directed along the third coordinate
axis, and the potential $A_0,$ both pointing in the third direction of
the group space. Namely,
\be
  \ol{A}^a_\mu=\delta_{\mu 2}\delta_{a 3}Hx_1+\delta_{\mu 0}\delta_{a 3}A_0=
  \delta_{a3}\ol{A}_\mu \,.
\ee{r12}
In order to account for the $A_0$-condensate, it is sufficient to
perform the substitutions $i\mu_1 \to gA_0$ and $i\mu_2 \to gA_0 /2.$

In the (2+1)-dimensional space-time, fermions are expressed in terms
of two-component spinors (see~(\ref{r2})), and thus they have no spin
degrees of freedom. Moreover, the Lorentz gauge fixing conditions
$a_0^b=0\,,\; D^ia_i^b=0$ lead to linear interdependence of the positive-
and negative-frequency solutions of the Lagrange equations
\be
  \left[ (\ol{D}^2)^{ab}\delta_{\mu\nu}+
  2g\ol{F}_{\mu\nu}^c f^{abc}\right]a^{b}_{\nu}=0 \;,
\ee{r13}
derived from (\ref{r5}). In fact, an arbitrary solution of
(\ref{r13}) can be represented as an eigenvector-series expansion:
\be
  a^\pm =a_1\pm ia_2= \sum_n N_n^\pm f_n(x),
\ee{r14}
where $N_n$ are constant coefficients, $f_n(x)$ are eigenvectors of
equation (\ref{r13}). The gauge fixing condition $D^-a^- + D^+a^+ =0$
can be shown to lead to the following restrictions, imposed on the
coefficients (see~\cite{trot}):
$N_n^-=2(n+2)N_{n+2}^+\:,\;n=0,1,2,\ldots;$ $\; N_1^+=0.$
No restrictions are imposed on the tachyonic mode coefficient $N_0^+.$
Thus, by setting chemical potentials $\mu_1$ and $\mu_2$ equal
to zero, we obtain the energy spectra of quark and gluon one-particle
excitations in the chromomagnetic field
\be
  \vep_\up{g}^2= 2gH(n-\frac{1}{2}) -i\epsilon, \ n=0,2,3,4, \ldots \,;
  \hspace{14mm}
\ee{r15}
\bear{c} \ds
  \vep_\up{q}^2= gHn +m^2 -i\epsilon,\quad n=0,1,2,3,\ldots \,,
  \quad\epsilon>0.
\eear{r16}
Infinitely small negative imaginary term $-i\epsilon$ gives a
prescription for handling the poles, as well as allows to determine a
correct limit for the tachyonic mode at $T\to 0$ (see (\ref{r23})).

\section{Gluon contribution to the free energy density} \label{sect}

In order to obtain an explicit expression for the effective potential,
one needs to substitute (\ref{r15}) and (\ref{r16}) in (\ref{r11}),
with account for the Landau levels degeneracy in a homogeneous
magnetic field. In the case of the (2+1)-dimensional
space-time this degeneracy is equal to $gHL^2/(2\pi).$ Then
the contribution of the charged gluon loop $v^\up{g}$ to the
thermodynamic potential $v$ can be represented in the form
\bear{rl}
  \ds v^\up{g}= \frac{gH}{2\pi\beta}&\ds \sum_{l=-\infty}^{+\infty} \left\{
  \ln\left[(\frac{2\pi l}{\beta}+gA_0)^2- gH -i\epsilon
  \right]+\right.\\ &\left.\ds\quad + \sum_{n=2}^\infty
  \ln\left[(\frac{2\pi l}{\beta}
  +gA_0 )^2 +2gH(n-\frac{1}{2}) -i\epsilon \right] \right\},
\eear{r17}
where the first term corresponds to the contribution of the tachyonic
mode (whose energy squared is negative). Limits of summation over
$l$ are infinite, and hence, expression (\ref{r17}) is periodic in $gA_0$
with the period equal to $2\pi/\beta.$ This fact is in close
relation with the gauge invariance property. When $T=0$, one can
perform the following transformation:
\be
  A_0^\prime=UA_0U^+ +\frac{i}{g}U\partial_0U^+ ,\; A_i^\prime=UA_iU^+,
  \quad \mbox{where } \; U=\exp\left(-igx_0A_0^a\frac{\lambda^a}{2}\right).
\ee{r18}
Then $A_0^\prime=0,$ and so $A_0$ is not a physical parameter. When $T\ne 0$,
the boundary conditions $A_\mu(x_0,{\rm\bf x})=A_\mu(x_0+\beta,{\rm\bf x})$
lead to subsidiary conditions $[U,\lambda^a]=0$ for all $\lambda^a.$ By
definition, it means that $U$ lies in the center of the gauge group.
Therefore, in the case of the $SU(2)$ group, only those gauge transformations
are allowed that preserve the $\mathbb{Z}_2$ symmetry:
\be
  A_0\to A_0^\prime=A_0+\frac{2\pi n}{\beta g},
  \quad \mbox{ where } n\in \mathbb{Z} .
\ee{r19}
The domain of gauge nonequivalent values of potential $A_0$ is given by
$gA_0 \in [0,2\pi T).$

It follows from (\ref{r17}) that $v^\up{g}$ is real, when the argument of
the first logarithm in (\ref{r17}) is positive, i.e. under the condition
\be
  \sqrt{gH} <gA_0 <2\pi T -\sqrt{gH},
\ee{r20}
otherwise the vacuum is unstable. As calculations demonstrate, the one-loop
effective potential in the (3+1)-case has a finite nontrivial minimum, which
however proves to be unstable~\cite{td1,td2}. According to~\cite{skalnew},
consideration for higher loops (ring diagrams) also leads to appearance of
an unstable minimum in the limit of high temperatures for fields of the
order of $(gH)^{1/2} \sim g^{4/3}T,$ which in the case of small
$\alpha_\up{s}$ exceeds the one-loop estimates for the field. Returning to
our general formula (\ref{r17}), we see, that under the condition
$gA_0=\sqrt{gH} $, the free energy $v^\up{g}$ becomes negative infinite. Such
singularity does not arise in the (3+1)-dimensional space, because of the
smoothing effect of integration over the momentum third component, absent in
the (2+1)-dimensional case.

In our case, the divergence can be eliminated by accounting for radiative
corrections, i.e. due to a nonvanishing imaginary part of the gluon
polarization operator (PO). We are interested in the energy radiative shift
for the tachyonic mode only, as it is responsible for the singular behavior
of the effective potential. In order to make qualitative estimates
without any detailed calculations, we write down the gluon PO in the form:
$\Pi(\vep,T) =\alpha_\up{s}\vep (\Pi_1+i\Pi_2),$ where $\Pi_1$
and $\Pi_2$ are some functions of the field and temperature, whose explicit
form has no essential influence on the further considerations and qualitative
results. Note, that $\Pi_2$ accounts for the gluon decay.
Contribution of the PO is of principle significance only in the vicinity of
the effective potential singularity. Consider the dispersion equation
$\vep^2=\vep_\up{g}^2+\langle\Pi(\vep, T)\rangle,$ and
make a reasonable assumption, justified by further numerical calculations,
that the behavior of the free energy near the singular point only
weakly depends on the values of the functions $\Pi_1\sim \Pi_2\sim1.$ Then,
the energy squared of the tachyonic mode, for $\alpha_\up{s} < (gH)^{1/2},$
with account for the radiative corrections and for the equality
$\varepsilon_\up{g}^2=-gH,$ can be approximately found to be:
\be
  \vep_\up{tach}^2 \simeq -gH -iC\alpha_\up{s} \sqrt{gH},
\ee{r21}
where $C$ is a coefficient of the order unity. As it is seen, the
nonvanishing imaginary part of $ \vep_\up{tach}^2$ guarantees the nonzero
argument of the first logarithm in~(\ref{r17}) at $gH\ne 0,$ though its
imaginary part is also nonvanishing.

Applying the known identity~\cite{prud}
\be
  \prod_l \left[1+\left(\frac{x}{2\pi l-a}\right)^2\right]=
  \frac{\op{ch}(x)-\cos(a)}{1-\cos(a)}
\ee{r22}
to (\ref{r17}) and discarding an irrelevant additive constant, similarly
to~\cite{td2}, one obtains
\bear{rl}
  \ds v^\up{tach}&\ds =\frac{gH}{2\pi}\left\{ \omega_0 +\frac{1}{\beta}
  \ln\left[1+\op{e}^{-2\beta\omega_0}-2\op{e}^{-\beta\omega_0}\cos(\beta gA_0)
  \right]\right\} = \vspace{3mm}\\
  &\ds =\frac{gH}{2\pi\beta} \ln\left\{ 2\cos\left[
  \beta(\sqrt{gH}+\frac{i}{2}C\alpha_\up{s})\right] -2\cos(\beta gA_0) \right\},
\eear{r23}
\be
  v^\up{g}=v^\up{tach}+ \frac{gH}{2\pi}\sum_{n=2}^{\infty}\left\{
  \omega_n +\frac{1}{\beta}\ln\left[ 1+\op{e}^{-2\beta\omega_n}
  -2\op{e}^{-\beta\omega_n}\cos(\beta gA_0) \right]\right\},
\ee{r24}
where
\be
  \omega_0=-i\sqrt{gH}+\frac{1}{2}C\alpha_\up{s}, \quad \omega_n^2= 2gH(n-\frac{1}{2}).
\ee{r25}

An expression for $v^\up{g}$ can be received with the help of the
Fock-Schwinger method. It is useful in studying the effective potential
in the region of small $gH.$ As the first step, recall the standard integral
representation
\be
  \ln A=-\int\limits_0^\infty \frac{ds}{s}\exp(-sA),
\ee{r26}
valid up to an additive infinite constant, and  apply it to the second
term in (\ref{r17}):
\be
  v^\up{g}=v^\up{tach}-\frac{gH}{2\pi\beta}\sum_{n=2}^\infty
  \sum_{l=-\infty}^{+\infty} \int\limits_0^\infty \frac{ds}{s}
  \exp\left\{-s\left[(\frac{2\pi l}{\beta}+gA_0 )^2 +2gH(n-\frac{1}{2})
  \right]\right\}.
\ee{r27}
Then, let us separate the function of temperature $v^\up{g}_T$ in the
effective potential $v^\up{g}=v^\up{g}_{T=0}+v^\up{g}_T$ (the zero
temperature part $v^\up{g}_{T=0}$ is independent of $A_0$ and will be
considered below, see (\ref{r35})). To this end, we transform the
summation over the Matsubara frequencies in (\ref{r27}) with the help of the
following identity (see~\cite{td1} or~\cite{prud}):
\be
  \sum_{l=-\infty}^{+\infty}\exp\left[-s(\frac{2\pi l}{\beta}+gA_0)^2\right]
  =\frac{\beta}{2\sqrt{\pi s}} \sum_{l=-\infty}^\infty
  \exp(-\frac{\beta^2 l^2}{4s}) \cos(\beta gA_0 l).
\ee{r28}
After performing summation over $n$ in (\ref{r27}) we obtain the final result:
\be v^\up{g}_T =
  v^\up{tach}_T-\frac{gH}{2\pi^{3/2}}\int\limits_0^\infty \!\!
  \frac{ds}{s^{3/2}} \left[\frac{1}{2\op{sh}(sgH)}-\op{e}^{-sgH} \right]
  \!\! \sum_{l=1}^\infty \exp(-\frac{\beta^2l^2}{4s}) \cos(\beta gA_0 l).
\ee{r29}
Numerical estimations confirm that the above expression (\ref{r29})
coincides with the temperature part of (\ref{r24}). The advantage of
expression (\ref{r29}) is that it has the evident limit at $gH\to 0$:
\be
  v^\up{g}_{T,H=0}=
  -\frac{1}{\pi\beta^3}\sum_{l=1}^\infty \frac{\cos(\beta gA_0 l)}{l^3}.
\ee{r30}
In the high temperature limit, $T\gg gA_0$, the following approximate
expression, demonstrating the nonanalytical dependence of the effective
potential on the background field, can be obtained:
\be
  v^\up{g}_{T,H=0}\approx -\frac{1}{\pi\beta^3} \left\{ \zeta(3)
  +\frac{(\beta gA_0)^2}{2} \left[\ln(\beta gA_0)-\frac{3}{2} \right]\right\}.
\ee{r31}

By setting $A_0=0$ in (\ref{r30}) one obtains twice the effective potential
of the uncharged gluons ($v^\up{g}$ contains the contribution of
gluons of two opposite color charges)
\be
  2v^{\up{g}_0}=-\frac{\zeta (3)T^3}{\pi}.
\ee{r32}
As has to be expected, we have obtained an analogue of the Planck law for
the black body radiation in the (2+1)-dimensional space-time. The total free
energy density can be obtained by adding $v^{\up{g}_0}$ to (\ref{r24}).
However, $v^{\up{g}_0}$ is the function of temperature only, and is
independent of the condensate field, and, hence, it is of no interest for us.

In order to obtain numerical estimates of the results, it is convenient to
pass to dimensionless variables
\be
  x=\beta \sqrt{gH},\, y=\beta gA_0.
\ee{r32a}
At $T=0$, this substitution is evidently incorrect. Nevertheless, in order to
make our notations universal, we will employ $x$ and $y$ even at zero
temperature, assuming that in this case $\beta$ takes on a certain finite
numerical value in the definition (\ref{r32a}). Let us add a constant to $v$
such that the dimensionless effective potential $u$ vanishes when the
condensate field goes to zero:
\be
 u(x,y,T)=\frac{v(H,A_0,T)}{T^3},\quad u(0,0)=0.
\ee{r33}
We have to find the minimum of the real part of the function
\be
  U(x,y,T)=U^{(0)}(x,T)+u(x,y,T),
  \quad U^{(0)}(x,T)=\frac{x^4}{2}\frac{T}{g^2},
\ee{r34} where the
quantity $T/g^2$ provides the scale of the temperature with respect to the
coupling constant $g.$
Along with the notations $u^\up{g}$ and $u^\up{q}$ for
the gluon and quark contributions to the dimensionless effective potential
$u$, let us also introduce the following one: $U^\up{g}=U^{(0)}+u^\up{g}.$
The zero-temperature part $u_{T=0}$ of potential $u$ is separated,
so that $u=u_{T=0}+u_T$ and $U^\up{g}_{T=0}=U^{(0)}+u^\up{g}_{T=0}.$

\begin{figure}
\epsfxsize=120mm \epsfysize=100mm
\centerline{\epsfbox{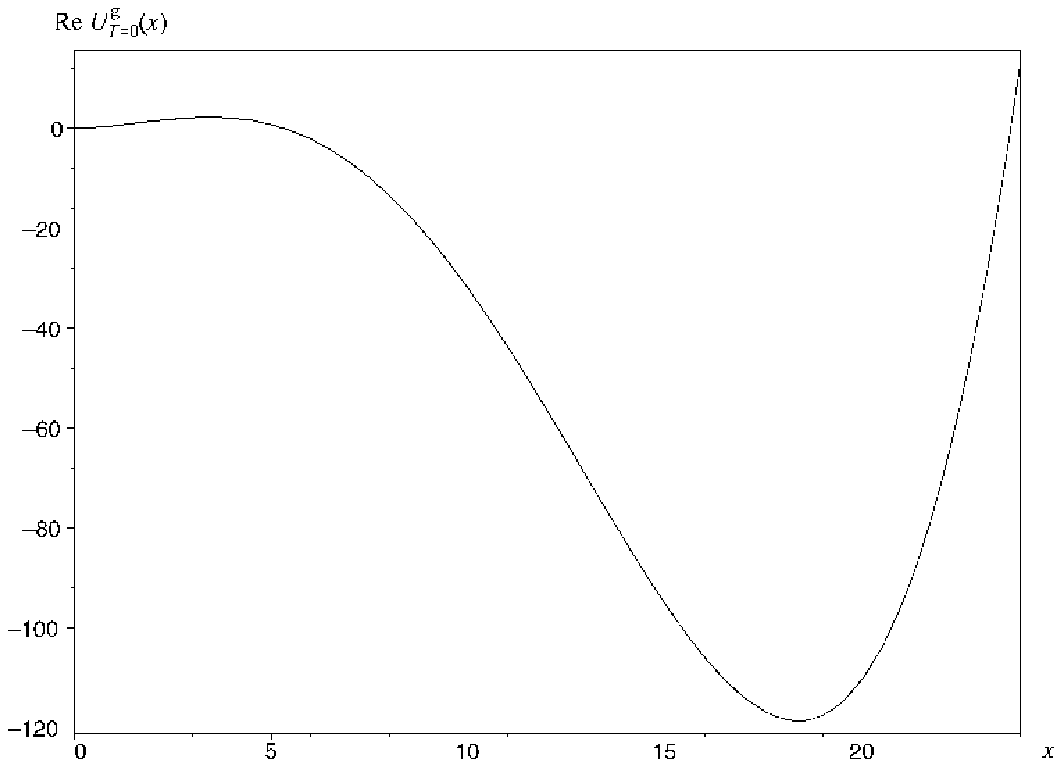}}
\caption{Бестемпературная часть $\op{Re} U^\up{g}_{T=0}(x)$ при $T/g^2=0.01.$}
\label{fig1} \end{figure}

Assuming $C=1$, the zero temperature contribution
($\beta\to\infty$) can be obtained from (\ref{r24}):
\be
  v^\up{g}_{T=0}=\frac{gH}{2\pi}\sum_{n=0,2}^{\infty}\omega_n=
  -\frac{(gH)^{3/2}}{2\pi}\left[1-\frac{\sqrt{2}-1}{4\pi}\zeta(\frac{3}{2})
  \right]-i\frac{(gH)^{3/2}}{2\pi}
  +\frac{gH}{4\pi}\alpha_\up{s}.
\ee{r35}
The real part of this expression coincides with the result obtained
in~\cite{trot} (see also~\cite{dittrich}). The contribution, provided
by the gluon tachyonic mode in the effective potential, was not obtained
in~\cite{trot}. Global minimum of $U^\up{g}$ (Fig.~\ref{fig1}) is
achieved at $\sqrt{gH} \approx 0.185 g^2.$ There is also a local minimum of
$U^\up{g}_{T=0}$ at $gH=0$ (in contrast to~\cite{trot}), which is formed by
the tachyonic mode contribution proportional to $gH$ and dominating
at $gH\to 0.$

The author of~\cite{trot} claims, that the condensate evaporates, when the
temperature is above some critical value $T>T_{\up{cr}}.$ The gluon tachyonic
mode was considered unphysical and omitted in that paper. However,
$v^\up{tach}$ gives nontrivial contributions both to the imaginary and real
parts of the free energy density at nonzero temperature, and hence, it has to
be taken into account. Moreover, it is the tachyonic mode that can generate a
minimum of $v^\up{g}$ at $A_0\ne 0$~\cite{td2}.

\begin{figure}
\epsfxsize=120mm \epsfysize=100mm
\centerline{\epsfbox{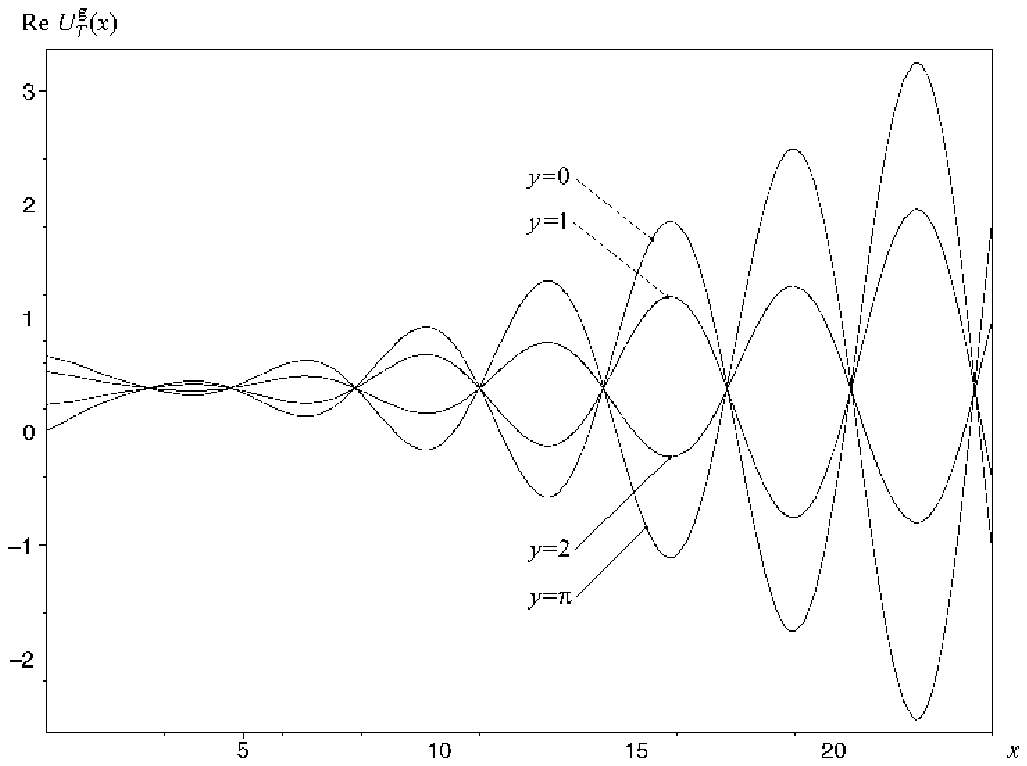}}
\caption{Температурная часть $\op{Re} U^\up{g}_{T}(x)$ при $T/g^2=0.01.$}
\label{fig2} \end{figure}

\begin{figure}
\epsfxsize=120mm \epsfysize=100mm
\centerline{\epsfbox{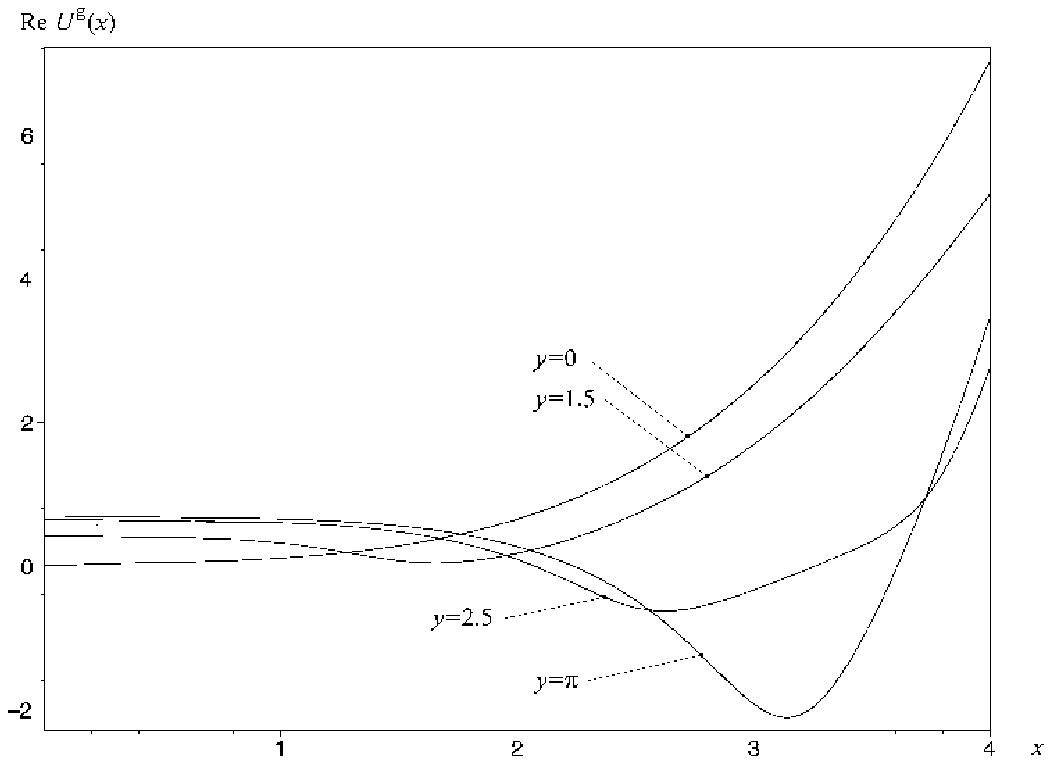}}
\caption{Потенциал глюонов $\op{Re} U^\up{g}(x)$ при $T/g^2=0.1.$}
\label{fig3} \end{figure}

\begin{figure}
\epsfxsize=120mm \epsfysize=100mm
\centerline{\epsfbox{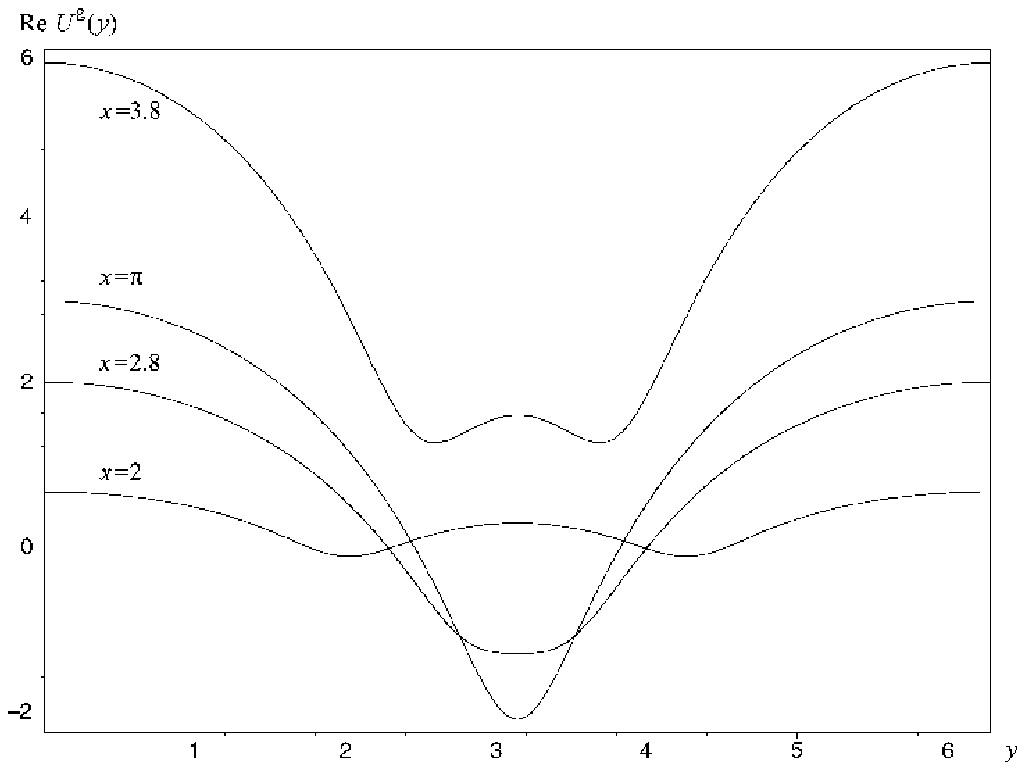}}
\caption{Потенциал глюонов $\op{Re} U^\up{g}(y)$ при $T/g^2=0.1.$}
\label{fig4} \end{figure}

Examination of the real part of $U^\up{g}(x,y=\op{const})$ as a function of
$x$ reveals a nontrivial minimum at
$x=x_{\min}$ that exists at the temperature $T$ lower than a certain critical
value $T_\up{cr}:\;$ $T<T_\up{cr}\sim 0.15 g^2,$ when the temperature
part $v^\up{g}_T$ is small with respect to the zero temperature
part $u^\up{g}_{T=0}.$ As an artefact of the normalization chosen,
the temperature contribution becomes weakly dependent on temperature,
while the zero temperature part becomes a function of temperature.
In this way, oscillating contribution $u^\up{g}_T$ (Fig.~\ref{fig2})
modulates the zero-temperature potential $U^\up{g}_{T=0},$ possessing a
nontrivial minimum (Fig.~\ref{fig1}). The value of $x_{\min}$ is close to
discrete points $n\pi, n\in\mathbb{N}$ and increases with growing temperature.
Moreover, a sequence of  second-order phase transitions between phases with
minima of $U^\up{g}(x_{\min},y)$ either at $y_{\min}=\pi$ or at $y_{\min}=0$
takes place. When temperature tends to zero, there appears an infinite number
of those phases. For example, at $T/g^2=0.1$, a global minimum of the
effective potential occurs for $x=3.03,\; y=\pi$
(Figs.~\ref{fig3},~\ref{fig4}). However, presence of an imaginary part of
$v^\up{g}$ means instability of this condensate configuration, at least at
the one-loop level (cf.~\cite{skalnew}). When the temperature increases above
$T_\up{cr},$ the condensate values $y_{\min}$ and $x_{\min}$ decrease from
$\pi$ to $0$, keeping correlation $x_{\min} \simeq y_{\min}$ unchanged.

As is well known, the system is in the confinement phase, if the trace of the
Polyakov loop~\cite{Polyak,Sussk} in the fundamental representation vanishes,
$\op{Tr}_\up{F}({\cal P})=0.$ The Polyakov loop is determined as follows:
\be
  {\cal P}= {\cal T}\exp\left[
  i\int\limits_0^\beta\!dt A_0^a\frac{\lambda^a}{2} \right].
\ee{r37}
In our case (\ref{r12}) potential $A_0$ is directed along the third axis in
the color space. Hence, $\op{Tr}_\up{F}({\cal P})= 2\cos(\beta gA_0/2)$.
It is evident, that the condition $\op{Tr}_\up{F}({\cal P})=0$ is satisfied,
when $\beta gA_0=\pi.$ Thus, the minima of the effective potential at
temperatures below the critical value corresponds to the confinement
and deconfinement phases.

\section{Quark contribution to the free energy density}

Let us consider the quark contribution to the free energy density in the same
manner, as it has been done for gluons in section~\ref{sect}. The quark
energy levels  degeneracy in the external chromomagnetic field is
proportional to the quark color charge, $\pm 1/2,$ and is equal to
$gHL^2/(4\pi),$ which is one half that of the gluon case. Upon substitution
(\ref{r16}) in (\ref{r11}), the expression for the quark and antiquark
effective potential is obtained:
\be
  N_\up{f}^{-1} v^\up{q}= -\frac{gH}{4\pi\beta}\sum_{l=-\infty}^{+\infty}
  \sum_{n=0,\lambda=\pm1\hspace{-5mm}}^\infty
  \ln\left\{\left[\frac{\pi (2l+1)}{\beta} +\frac{\lambda}{2}gA_0 \right]^2
  +gHn+m^2 \right\}.
\ee{r42}
Here $\lambda=\pm 1$ corresponds to the quark color charge values.
The essential difference from the nonabelian gauge field contribution
in the chromomagnetic background is the absence of the tachyonic mode
in the quark energy spectrum. Therefore, $v^\up{q}$ is a well defined
function of the condensate field in the whole domain
$0<gH<\infty,\;0<gA_0<\infty:$
\be
  N_\up{f}^{-1} v^\up{q}=-\frac{gH}{2\pi}\sum_{n=0}^\infty\left\{ \omega_n +
  \frac{1}{\beta}\ln\left[1+\op{e}^{-2\beta\omega_n}+2\op{e}^{-\beta\omega_n}
  \cos(\frac{\beta gA_0}{2})\right] \right\},
\ee{r43}
where
\be
  \omega_n^2=gHn+m^2.
\ee{r44}
The quark field at finite temperature is subject to the fermionic
antiperiodicity condition $\psi(x_0,{\rm\bf x})=-\psi(x_0+\beta,{\rm\bf x}).$
Therefore, the period $4\pi T$ of (\ref{r43}) in variable $gA_0$ is twice the
period of (\ref{r24}). This implies that the residual gauge $\mathbb{Z}_2$
symmetry is violated ($\mathbb{Z}_N$ in the case of the $SU(N)$ gauge group).
In what follows, we restrict ourselves to consideration of the quark field in
the chiral limit, $m=0$. In this case the zero temperature quark contribution
is equal to
\be
  N_\up{f}^{-1}v^\up{q}_{T=0}=-\frac{gH}{2\pi}\sum_{n=0}^\infty\sqrt{gHn}=
  \frac{(gH)^{3/2}}{8\pi^2}\zeta(\frac{3}{2}),
\ee{r45}
which coincides with the result of~\cite{trot}. In contrast to the
case of the gluon potential $v^\up{g}_{T=0},$ the sign of $v^\up{q}_{T=0}$ is
always positive, and this does not allow a nontrivial minimum to be
formed. The temperature dependent part of the quark effective
potential reads
\be
  N_\up{f}^{-1}v^\up{q}_T=
  -\frac{gH}{2\pi\beta}\sum_{n=0}^\infty
  \ln\left[1+\op{e}^{-2\beta\omega_n}+2\op{e}^{-\beta\omega_n}
  \cos(\frac{\beta gA_0}{2})\right].
\ee{r46}
By using the proper time method one can obtain an alternative
representation, which evidently demonstrates continuous behavior of
$v^\up{q}_T$ at $gH\to 0\,:$
\be
  N_\up{f}^{-1}v^\up{q}_T= \frac{gH}{2\pi^{3/2}}
  \int\limits_0^\infty\frac{ds}{s^{3/2}} \sum_{l=1}^\infty
  \exp(-\frac{\beta^2l^2}{4s})
  \frac{\cos(\beta gA_0l/2+\pi l)}{1-\op{e}^{-sgH}}.
\ee{r47}
In the limit $gH\to 0$ we obtain
\be
  N_\up{f}^{-1}v^\up{q}_{T,H=0}=\frac{2T^3}{\pi} \sum_{l=1}^\infty (-1)^l
  \frac{\cos(\beta gA_0l/2)}{l^3}.
\ee{r48}

\begin{figure}
\epsfxsize=120mm \epsfysize=100mm
\centerline{\epsfbox{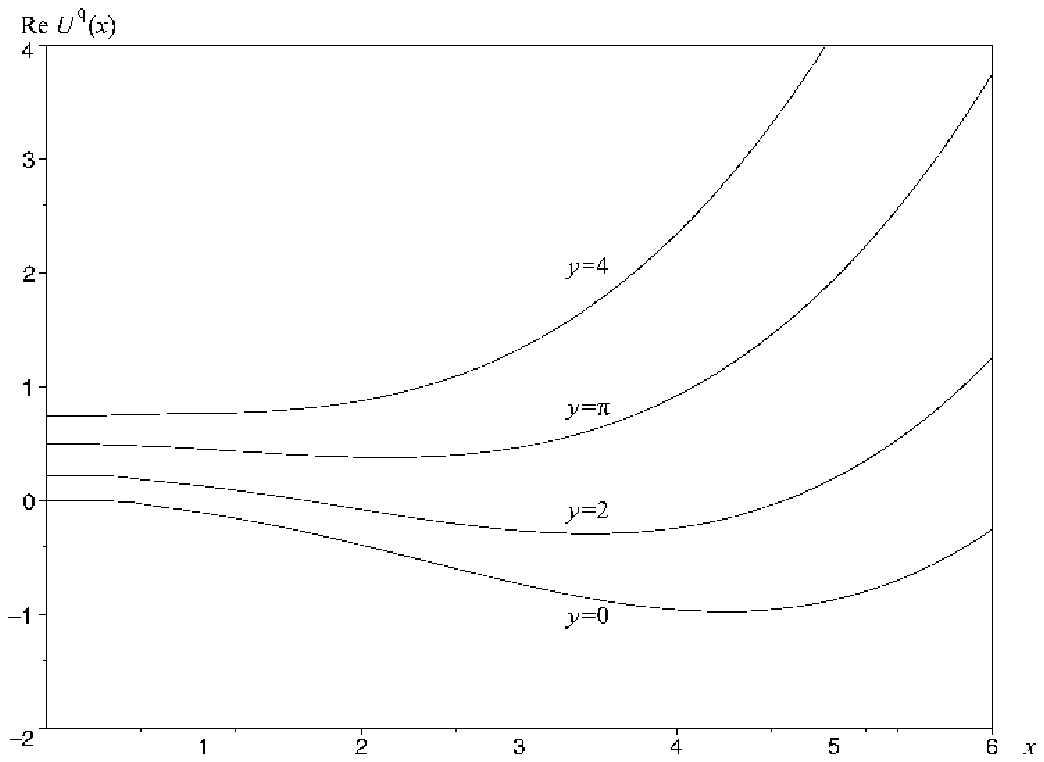}}
\caption{Вклад кварков $\op{Re} u^\up{q}(x).$}
\label{fig5} \end{figure}

\begin{figure}
\epsfxsize=120mm \epsfysize=100mm
\centerline{\epsfbox{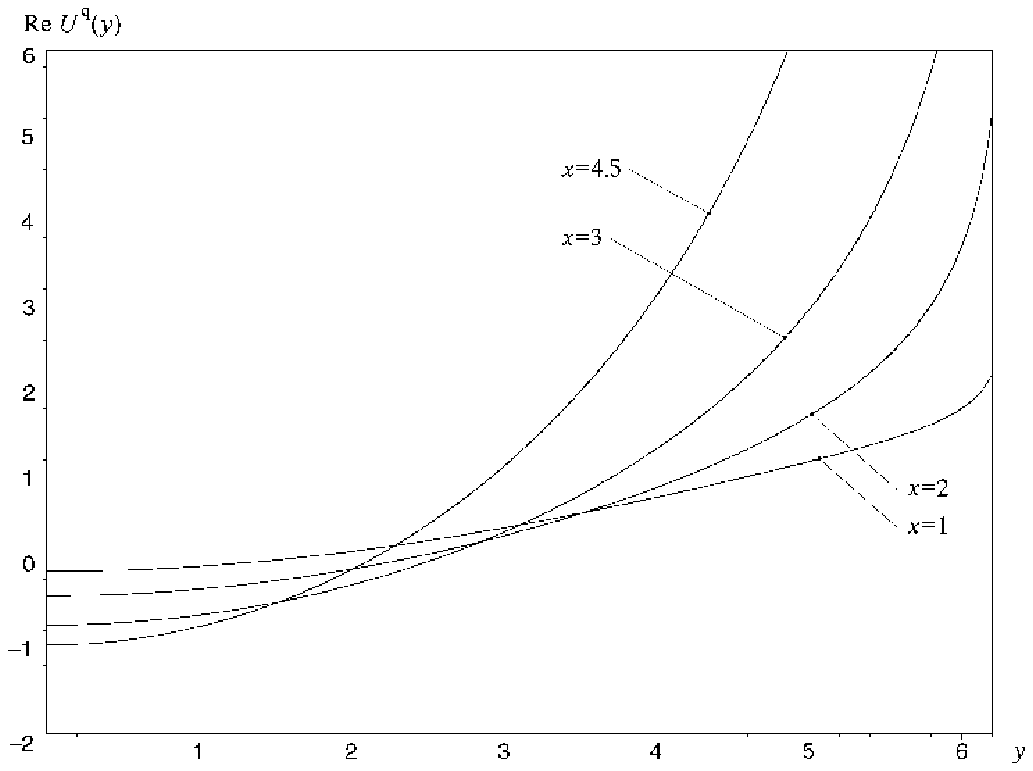}}
\caption{Вклад кварков $\op{Re} u^\up{q}(y).$}
\label{fig6}\end{figure}

The dimensionless effective potential $u^\up{q}(x,y)$, defined in (\ref{r33}),
does not depend on temperature $T.$ The family of curves for $u^\up{q}$
with fixed values $x=\op{const}$ and $y=\op{const}$ are depicted in
figs.~\ref{fig5},~\ref{fig6} respectively. It is evident that for any
$x=\op{const}$ the minimum of $u^\up{q}(x,y)$ is reached at $y_{\min}=0.$
Nevertheless, there is a nontrivial minimum at $y=\op{const}<3.5$ for
$x_{\min}\ne 0.$ Global minimum of the gluon effective potential
is reached at $x=4.30, y=0.$

The quark contribution to the free energy is greater than the gluon one
at $y>4,$ when the quark zero mode plays a dominant role. Therefore,
our conclusions about the existence of the field condensate and of the
confinement-deconfinement phase transitions at $0\le y\le \pi$ remain
unchanged even when the total free energy is considered.

\section*{Conclusions}
Thus, in the present paper, contributions of gluons and quarks to the
thermodynamic potential (free energy) in the (2+1)-dimensional space-time are
calculated in the one-loop approximation at finite temperature in the
background of the superposition of the uniform constant chromomagnetic field
$H$ and the $A_0$-condensate. Consideration of the role of the tachyonic
mode in the gluon energy spectrum leads us to a conclusion that it can not be
neglected, as opposed to what has been done in~\cite{trot}. Moreover,
accounting for the one-loop radiative correction to the gluon energy spectrum
enables a nonanalytic behavior of the effective potential, related to the
presence of the zero modes in the energy spectrum, to be cured. The free
energy minimum is studied and a possibility of its formation is demonstrated
at certain nonzero values of the chromomagnetic field strength $H$ and of the
potential $A_0.$ The analysis of the temperature dependence of the results
demonstrates that below a certain critical value of temperature a number of
subdomains in the domain of the parameters exist where the system is either
in the confinement or in the deconfinement phases. This behavior is explained
by the oscillating contribution of the tachyonic mode to the free energy
density. Unfortunately, the imaginary part of the effective potential does
not vanish at points $x_{\min}\simeq\pi n, n\in\mathbb{N},$ where $V(x,y)$
reaches its minimum. Hence the nontrivial minimum of the effective potential
generated by the condensate fields $A_0$ and $H$ turns out to be unstable.
This instability is related to the choice of the homogeneous vacuum field,
and it is reasonable to assume, that it may be removed in the realistic
situation of a vacuum field inhomogeneous at large distances, where
confinement is formed (see also the arguments of~\cite{td2}). With regard for
what has been said, the states found in the present paper may be considered
as quasistable. More rigorous justifications for the assumptions made above
might be obtained in the study of inhomogeneous vacuum field models (see,
e.g.~\cite{eng1}), and, following~\cite{skalnew}, in elaboration of higher
loop contributions.

\newpage

\end{document}